\documentclass[12pt,a4paper]{article}
\usepackage{graphicx}
\usepackage{cite}
\usepackage{float}
\usepackage{amsmath,amssymb,amsfonts}

\newcommand{\bse}{\begin{subequations}}
\newcommand{\ese}{\end{subequations}}
\newcommand{\be}{\begin{equation}}
\newcommand{\ee}{\end{equation}}
\newcommand{\bea}{\begin{eqnarray}}
\newcommand{\eea}{\end{eqnarray}}
\newcommand{\ba}{\begin{array}}
\newcommand{\ea}{\end{array}}

\def\ZZ{{\mathcal{Z}}}

\input amssym.def
\input amssym.tex

\usepackage[colorlinks=true, linkcolor=blue, bookmarks=true, citecolor=red]{hyperref}

\usepackage{geometry}
\begin{document}

\title{Evaporation/Hadronization Correspondence}
\date{}

\maketitle

\hrule

\begin{center}

\author{Davood Allahbakhshi\footnote{allahbakhshi@ipm.ir}\\\vspace{5mm} School of Particles and Accelerators, Institute for Research in Fundamental Sciences (IPM), P.O.Box 19395-5531, Tehran, Iran}
\end{center}

\hrule

\linespread{1.5}

\abstract{
\noindent
A holographic duality is proposed between black hole evaporation in the bulk and hadronization (confinement) in dual field theory. Information paradox is discussed in this duality. We also propose that the recently introduced semi black brane solution is holographically dual to a mixed plasma of quarks, gluons and hadrons in global equilibrium.}

\newpage

\hrule

\tableofcontents

\vspace{5mm}
\hrule

\section{Introduction}

In 1950 Fermi presented the idea of using statistical concepts to predict the final state of a fireball \cite{Fermi:1950jd}, a very hot and dense lump of partons. Later during 1960’s Hagedorn developed this idea \cite{Hagedorn:1965st,Hagedorn:1967ua,Hagedorn:1968jf}. He assumed a self-similarity between fireballs. The first exciting result was the appearance of an exponential growth of the density of states of the fireball as the mass of the fireball increases. This result led immediately to appearance of the Hagedorn temperature which is the highest achievable temperature by the system. During 1970’s and 1980’s physicists could explain the momentum distribution of final hadrons by using relativistic kinetic theory, Hydrodynamical models and Cooper-Frye equation \cite{Cooper:1974mv} simultaneously. During 1970’s, 80’s and 90’s we found experimental signs that the system which is produced at high enough temperatures ($\sim$ 175 Mev) is in fact a compressible fluid instead of a free gas. By starting the RHIC and LHC experiments, the heavy ion collision experiments entered a new range of energy and accuracy.

On the other hand the AdS/CFT conjecture \cite{Maldacena:1997re,Witten:1998qj}, increased hopes for studying strongly coupled field theories by just studying a dual classical gravitational model. From the first days after Maldacena's conjecture, QCD was one of the first theories under consideration. In 1998 Witten proposed that the holographic dual to a field theory in a thermal state is a black hole in the bulk \cite{Witten:1998zw}, which enabled us to make holographic duals for thermal systems including quark-gluon plasma. Many attempts started to make holographic models for QCD. The most successful ones were Sakai-Sugimoto \cite{Sakai:2004cn}, Hardwall and Softwall models \cite{Erlich:2005qh,Karch:2006pv}. Later Kiritsis and Gursoy introduced another model, IHQCD \cite{Gursoy:2007cb,Gursoy:2007er}.

There are also many QCD-related concepts and quantities that have been studied in many different situations and backgrounds including fluid/gravity duality \cite{Bhattacharyya:2008jc}, the drag force \cite{Gubser:2006bz,AliAkbari:2009pf,Abbasi:2012qz,Abbasi:2013mwa}, QCD phase space and its critical points \cite{DeWolfe:2010he}, hadron spectrum \cite{Karch:2006pv}, thermalization \cite{Danielsson:1999fa,Ebrahim:2010ra}, meson melting \cite{Peeters:2006iu,Hoyos:2006gb,Ali-Akbari:2014xea} and many more.

Although the confinement/deconfinement transition is also studied in gauge/gravity duality \cite{Herzog:2006ra} but, there is an obvious lack of a comprehensive \emph{dynamical} holographic description for the hadronization process, consistent with all other mentioned studies in AdS/QCD literature such as holographic hydrodynamics and thermalization models. In present paper we propose such a holographic dual.

Any dual picture for black hole evaporation should naturally lead to some explanation for the black hole information paradox \cite{Hawking:1976ra}. The paradox arises when we use the quantum field theory for matter fields on a classical background geometry which includes a horizon. From calculations by Hawking \cite{Hawking:1974sw} we know that in such cases a thermal beam of radiated particles will appear if we ignore the backreaction of the radiation on the geometry. Ignoring this backreaction seems not to be completely consistent with equivalence principle, unitarity and locality. The result is the information paradox which can be expressed in different ways and also produce other problems such as the problem of firewall \cite{Almheiri:2012rt}.

Many different ideas are proposed for solving the paradox but it seems that we have just two ways to solve the problem. Showing that including the backreaction of the radiation is important enough to solve the problem or giving up one of the accepted principles of general relativity or quantum mechanics which can potentially lead to other difficulties.

In present paper we propose a duality between black hole evaporation in the bulk of the AdS space and \emph{collective} hadronization (confinement) in dual field theory. Different sides of the duality is discussed and a dictionary is presented for relating concepts in both processes. We present some ideas related to information paradox in this holographic setup. We also propose that the recently found semi black brane solution \cite{Allahbakhshi:2016gyj} corresponds to a mixed plasma of quarks, gluons and hadrons in global equilibrium.

The paper is organized as follows. At first we explain the proposal and the evidences that support it. Then we discuss the information paradox in this holographic picture. Finally holography of the semi black brane is discussed.

\section{Evaporation/Hadronization Correspondence}
In this section we mention to some evidences which lead us to the proposal that the AdS black hole evaporation is dual to hadronization (confinement) in dual field theory. For this we need to have a brief review on some theoretical attempts to explain the process of hadronization specially the statistical model of hadronization. We then explain the proposal and its physical implications. And finally we will have a discussion on information paradox.

\subsection{Statistical Model of Strong Interactions}
As mentioned, in 1950 Fremi proposed a statistical model for hadronization \cite{Fermi:1950jd}. Hagedorn developed this work more, during 1960s \cite{Hagedorn:1965st,Hagedorn:1967ua,Hagedorn:1968jf}. The main unknown quantity was the density of states of a fireball of mass $m$. Hagedorn supposed that massive enough fireballs are \emph{self similar} and he assumed that \emph{fireballs are made of fireballs.} This assumption implies that the density of states of a fireball should be equal to the multiplication of the densities of all smaller fireballs which can make the original fireball.
\be
\rho(m)=\sum_{n=2}^{N}\prod_{i=1}^{n}\rho(m_i)\delta(\sum_i m_i-m).
\ee
It is obvious that the solution to this equation should be an exponential function. Later in 1972 Nahm found an exact analytic solution to this equation \cite{Nahm:1972zc}.
\be
\rho(m)=const. \; m^{-3}e^{m/T_H},
\ee
where $T_H$, named \emph{Hagedorn Temperature}, is a constant related to some quantities such as volume and mass of the lightest fireball. If we use this density to calculate the partition function of the fireball it is something like
\be
ln \ZZ = V \left( \frac{T}{2\pi} \right)^{3/2}\int dm\;m^{-3/2}e^{-m(1/T-1/T_H)},
\ee
which means that there should be some phase transition at $T_H$. Later we found that it is the phase transition \emph{from hadron gas to quark-gluon plasma}.

\begin{figure}[H]
\centering
\includegraphics[scale=1]{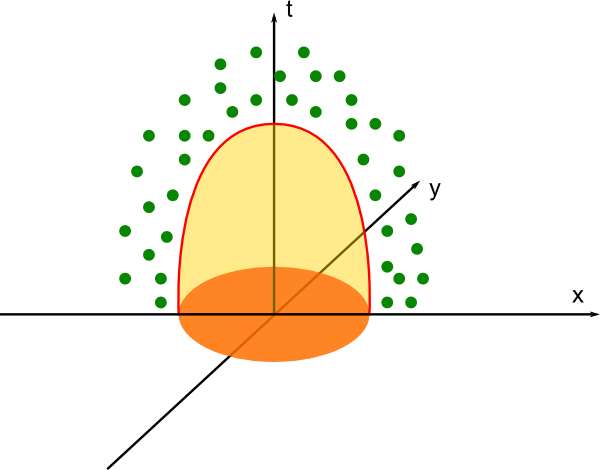}
\caption{\label{shockpic}The sketch of shock front in space-time.}
\end{figure}

There are also some other theoretical tools to study the hadronization process. One of the most interesting ones is the shockwave model. In this model the system is considered as a non-normal relativistic shock wave. There is a \emph{shock front hypersurface} named \emph{freezout hypersurface} in space-time. On one side of this hypersurface we have quark-gluon plasma and on the other side we have the hadron gas. A sketch of such hypersurface is shown in fig. \ref{shockpic}. Although the front is considered as a mathematical hypersurface but it has a thickness in reality in which the system is in the mixed phase, a mixture of quarks, gluons and hadrons.
It is known that at final moments of hadronization, the freezout hypersurface becomes space-like which means that the fireball will be hadronized in a \emph{non-causal} way \cite{Glendenning:1984fi,Kajantie:1986cu}. The physics of this event is so simple. During hadronization the fireball becomes cool and dilute. At some moment the plasma becomes \emph{supercooled} which means that the energy density of the plasma becomes smaller than the energy density of the nuclear matter ($0.134 \; Gev/fm^3$). Beyond this point the hadron gas is \emph{thermodynamically favored} in comparison to quark-gluon plasma and so rest of the plasma will be hadronized approximately at the same moment, non-causaly. This event is very important and we will discuss it in the holographic picture.

\subsection{Evaporation/Hadronization Correspondence: the proposal}

The AdS/CFT correspondence \cite{Maldacena:1997re,Witten:1998qj} is a very interesting and in some cases powerful tool to study strongly coupled quantum field theories. According to the AdS/CFT dictionary the holographic dual to a field theory in a thermal state, is an AdS black hole in the bulk \cite{Witten:1998zw}. For example from the behaviour of the Wilson loop we know that the system is in the deconfined phase\footnote{Some physicists are not sure that this interpretation is completely correct. For example you can have a look at \cite{Mandal:2011ws}. But most of physicists believe that the holographic dual to a system in deconfined phase is a black hole and even there are some firm evidences for it \cite{Hanada:2014noa,Hanada:2014xya}. In this paper we also use the latter idea.}. It means that we can start from pure AdS geometry and then inject some energy/matter into the bulk. Since there is a background gravitational force from boundary to the origin of the AdS geometry and also since the gravity is a non-linear force, the injected energy will be concentrated at the origin of the space after a while and a black hole forms. The black hole formation has in fact two steps. The first one is the appearence of the horizon and the second step is relaxation and equilibration of the horizon locally and finally globally. As soon as the horizon is formed the Wilson loop tells us that the system is deconfined. Relaxation and equilibration of the horizon tells that the dual system is equilibrated. So it is very natural to consider the black hole formation as the holographic dual to formation and thermalization of the quark-gluon plasma in dual QCD-like field theory. It is the standard picture of thermalization in holography \cite{Danielsson:1999fa}. Then we can study many things in this black hole geometry. One of the very important ones is fluid/gravity correspondence \cite{Bhattacharyya:2008jc}, which states that we can exactly find the boundary hydrodynamics, order by order, by starting from a black brane. We can also calculate some transport coefficients of the quark-gluon plasma from this fluid/gravity correspondence as well as some other ways such as \emph{linear response theory}. In all these holographic models the black hole corresponds to deconfined matter.

On the other hand, perturbative modes of matter fields out of the black hole can be considered as some bound states in dual field theory. For example in AdS/QCD models such as Hard Wall \cite{Erlich:2005qh} and Soft Wall \cite{Karch:2006pv} models, the spectrum of the normal modes of these fields have been considered as the spectrum of the Hadrons in QCD and the agreement is also surprising. The Regge trajectories can even be made by appropriate backgrounds, e.g. the backgrounds of the Soft Wall model or IHQCD.

When there is a horizon in the space, we have not the stable normal modes anymore. In contrast we have unstable \emph{quasi-normal modes}, correspond to some unstable bound states in a hot quark-gluon plasma. Characteristics of these modes can be assigned to their counterparts in dual theory. The energy of these modes corresponds to the energy (mass) of these bound states and the life-time of the modes will be the life-time of the unstable bound states in dual theory. The black hole eats these modes up. It means that the unstable bound states are being melted into the plasma \cite{Hoyos:2006gb}, since the black hole represents the deconfined matter (quark-gluon plasma) and the modes of matter fields, around the black hole, represent the confined matter (hadrons) in dual theory.

There is an important point. If there are many such matter modes around the black hole, then these modes altogether can not be considered as fluctuations of the field and in fact they produce a classical background matter field to make the black hole hairy, which means that the vacuum state of the dual theory is changed. The question here is this: \emph{What is the dual to a hairy black hole in an AdS/QCD model?}

The answer to this question is obvious. Coexistence of the horizon and the matter field hair, means that we have quark-gluon plasma and hadron gas in a thermal dual theory simultaneously. So \emph{the dual system to a hairy black hole in an AdS/QCD model is a QCD system in a mixed phase\footnote{But in the saddle point approximation, in the language of path integral.}.}

Now suppose that this hair is produced by the Hawking radiation. It immediately means that the Hawking radiation is responsible for the production and condensation of hadrons in dual theory. The point that the Hawking radiation is dual to hadron production in dual theory can be understood in many other ways. For example any radiated Hawking  quanta corresponds to appearence of a bound state, e.g. a meson, from the quark-gluon plasma in dual system. Any eaten (quasi-normal) mode by the black hole, corresponds to a melted hadron into the plasma.

When the black hole is evaporated completely, there is not any horizon in the bulk, but just a matter hair, means that the system is in the confined phase. The dual picture is also correct. The quark-gluon plasma has changed it's phase to the confined matter (hadrons) completely. This is nothing but the \emph{hadronization process}.

The holographic picture should be complete, means that anything at one side should have a holographic counterpart on the other side. Here we propose a dictionary for all these things:
\begin{itemize}
\item \emph{\textbf{Hadrons:}} As mentioned the matter modes around the black hole  can be identified with the hadrons in dual theory. A matter hair around the black hole corresponds to a huge number of hadrons in dual system and the state of the system is not a simple quark-gluon plasma anymore, but a mixed phase.

\item \emph{\textbf{Thermality:}} One of the immediate results from the proposed duality between evaporation and hadronization is the thermal spectrum for all the produced hadron species. Since the Hawking radiation is thermal, the produced hadrons should be thermal as well, at least as long as we do not consider the backreaction of the radiation on the geometry. But this is an old wellknown fact, we know from \emph{Heavy Ion Collision} experiments that the spectrum of the produced hadrons is thermal for all hadron species\footnote{In some special cases the spectrum is near thermal. For example the spectrum of pions at low transverse momentums shows a deviation from thermality, well known as \emph{pion spectra enhancement at low transverse momentums}. There are many different proposals for explaining these deviations. Just as first examples you can have a look at \cite{Gavin:1991ki,Ornik:1991dm} and their citations. It needs investigation that our holographic proposal can explain this non-thermality or not, if we consider some non-thermal effects in the Hawking radiation, due to backreactions.}. This is one of the interesting experimental facts that how such a fast decaying plasma can be thermally equilibrated.

\item \emph{\textbf{Greybody Factor:}} When the horizon radiates particles, some of them are reflected back to the horizon by the curved space and the final spectrum of the particles at the boundary is not completely that of a black body but a grey body. This portion of the radiated particles which reach the boundary is named greybody factor. It depends on the geometry and also the frequency of the radiated particle. In our holographic picture this greybody factor has an obvious meaning. \emph{Not all produced hadrons can survive and remain stable in the hot quark-gluon plasma}. Some of them will be melted back into the plasma and the final spectrum of hadrons is not that of a black body but a grey body. This effect should be related to \emph{quarkonium suppresion} in dual system, in the sense that the grey body factor should be smaller for more energetic modes. This is in fact the case! for example it is shown that the greybody factor for large frequencies in AdS geometry behaves like $\Gamma(\omega)\sim A/\omega^{d-1}$ for a $d+1$-dimensional black hole \cite{Harmark:2007jy}, where $A$ is the area of the black hole which is related to the entropy of the system. This result shows that the greybody factor decreases by increasing the energy of the mode; so may be consistent with quarkonium suppression. It is also shown that the melting time of the mesons decreases by increasing the temperature and anisotropy of the plasma \cite{Hoyos:2006gb,Ali-Akbari:2014xea}.

\item \emph{\textbf{Hadronization Time:}} If the black hole evaporates completely, the evaporation time of the black hole should be the hadronization time of the quark-gluon plasma.
\item \emph{\textbf{Hadron Ratios:}} If there are more than one matter filed in the model, then the black hole  will evaporate to them all. The ratios of different radiated particles should be consistent with \emph{hadron ratios} which is known from heavy ion collision experiments. Of course it needs more investigations.

\item \emph{\textbf{Equilibration:}} AdS black holes can be equilibrated with their radiation. This happens when the AdS black hole  is very large. It means that a dense enough plasma can be equilibrated with its produced hadrons finally and the final phase of the system is a mixed phase. Note that in the holographic model under consideration, a black hole corresponds to a plasma which fills whole the space uniformly.

\item \emph{\textbf{Hawking-Page Transition:}} As mentioned previously, if the plasma is not dense enough it will experience a non-causal hadronization process, at final moments of its life, because of the supercooling effect. At this moment the hadron gas is thermodynamically favored in comparison to the quark-gluon plasma. We propose that this sudden hadronization of the plasma should correspond to a real time Hawking-Page-like phase transition where the pure AdS geometry is thermodynamically favored in comparison to the black hole. A large AdS black hole\footnote{But not large enough to be equilibrated with its radiation. By calculating the energy-momentum tensor of the plasma from holography we easily see that such a black hole corresponds to a not dense enough plasma.}, will be suddenly replaced by a pure AdS-like geometry. Note that this interpretation is consistent with the usual interpretation of the Hawking-Page phase transition, as a confinement/deconfinement phase transition in holography \cite{Herzog:2006ra}.

\item \emph{\textbf{Information Paradox:}} When we speak about the black hole  evaporation we can not ignore the problem of information paradox. This is probably the most interesting and challenging part of the proposal. We will discuss the information paradox in this holographic setup.

\end{itemize}

\subsection{Information Paradox and Holography}
From the first months after Maldacena conjecture, it was known that if the correspondence is correct, the information paradox should be solved somehow in this context. Since any evolution in field theory is unitary, its holographic dual should be unitary as well. So if the black hole evaporation has any counterpart in dual field theory, it should be unitary and we should be able to explain about the fate of the information after evaporation of the black hole.

Information paradox can be expressed in this way: suppose that we prepare some matter in a pure state. This matter collapses and becomes a black hole. Black hole evaporates thermally and finally we have just the radiated matter in a thermal state, which is mixed. The problem is that it is not possible to move from a pure state to a mixed state by unitary evolution. It seems that there are just two possible resolutions. Including \emph{non-unitary} evolutions in quantum mechanics, which seems too bad! The other resolution is deviation from thermality, somehow, which is needed to encode the information of the initial matter into the final radiation. Most physicists think that the second resolution is probably correct because of some points. The first, and probably most important one, is that the Hawking radiation is thermal because the backreaction of the radiation on the black hole is ignored. If we consider the effect of backreaction then the radiation is not thermal anymore. But the question is that is this non-thermal deviation large enough to carry the huge amount of the initial information in it? some physicists think that this deviation is so small \cite{Mathur:2009hf} and some of them argue that it can be large enough \cite{Dvali:2015aja}. Some people try to describe the paradox by the hyothesis of \emph{black hole complementarity} \cite{'tHooft:1990fr,Susskind:1993if}. There are some other paradoxes related to information paradox such as \emph{firewall paradox} \cite{Almheiri:2012rt} which states that it is not possible to keep equivalence principle, unitarity and locality simultaneously.

In present context, if there is any correspondence between black hole evaporation and hadronization in field theory, we should be able to explain about the information from the hadronization process. Let us have a look on what we have on both sides of the process

- On gravity side we have a mass/energy injection into the space. On field theory side it corresponds to energy injection into the hadronic system, e.g. by colliding lead nuclei.

- On gravity side we have black hole formation and relaxation. On field theory side we have quark-gluon plasma formation and relaxation.

- On gravity side we have black hole evaporation. On field theory side we have quark-gluon plasma hadronization.

- If the black hole is not very large, on gravity side we have complete evaporation and just a matter hair in the final state. On filed theory side we have a hadron gas as the final state of the system.

The information paradox in the language of hadronization should be something like this: We start from two nuclei in a definite pure state. They collide and change their phase to quark-gluon plasma. The produced fireball hadronizes which seems to be thermally and finaly we just have some hadrons ($\sim 10,000$ hadrons in LHC). \emph{Where is the information of the initial nuclei? And is the final state pure?}
There are some possibilities:

1 - The hadronization process is not completely thermal and there is a small deviation from thermality which is sufficient to carry the information of the initial nuclei.

2 - The information will be released during the final, non-causal decay of the plasma, because of supercooling. This sudden, non-causal hadronization, does not seem to be thermal.

3 - The key is in entanglement not just naive (non-)thermality. The hadronization process has two periods. The causal period and the non-causal period. during the causal period any radiated hadron from the plasma (Bob hadron) is \emph{entangled} with a counterpart in the plasma (Alice hadron), which is not necessarily a bound state. This is also correct for all radiated Bob hadrons during causal period. Finally all Alice partons will be released during the non-causal period of the hadronization. The information is encoded in all entangled hadrons, in a very complicated way.

4 - When the hadronic matter changes phase to \emph{quark-gluon plasma}, the information of the initial \emph{hadronic matter} will be lost (!) since the information of the initial hadrons is related to some effective objects (hadrons) and they can not be encoded in quarks and gluons. In other language the initial state of partons is not known. In this way we can just read the conserved charges of the initial state in the final state, nothing more.

All these possibilities have obvious holographic duals. Probably it is possible to add more cases to this list and having any definite opinion about them needs more investigations. The only implication of the current proposal is that any explanation about the information during a heavy ion collision should be also applicable to black holes, at least partially.

\subsection{Holographic Interpretation of Semi Black Brane}
Although it needs investigations but, we can speculate about the holography of the semi black brane solution which is introduced in \cite{Allahbakhshi:2016gyj}. This geometry is asymptotically AdS. The geometry has some features of the black brane and pure AdS space simultaneously. It does not have a horizon so the dual system should not be completely deconfined, in the language of AdS/QCD, and on the other hand penetrating deep into the space from the boundary is hard, so the dual system is not completely confined. It is interesting to study \emph{Willson loop} and \emph{meson dissociation} in this geometry to find the potential between a quark and anti-quark. Also it seems that we can not assign any concept of temperature to this solution but, there is a parameter $\rho_h$ which will appear in the boundary energy-momentum tensor. This parameter in combination with the coupling $C_1$, should be related to the energy density and/or pressure in dual field theory. There is not any horizon so we do not expect any Hawking radiation and also any quasi-normal mode to exist, means that there is not any net \emph{hadron production} and \emph{melting} in dual theory, but just some stable hadrons corresponding to \emph{normal modes} in the bulk, \emph{effectively}. It will be interesting to study the spectrum of these modes. From all these evidences we propose that the semi black brane corresponds to \emph{a system of quarks, gluons and hadrons in an equilibrated mixed phase.} Since the semi black brane is completely planar, the mixed plasma is in global equilibrium. The rates of meson production and melting are the same and cancel each other, means that there should not exist any horizon in the bulk \emph{effectively}. The system is neither confined nor deconfined! In this way it is natural to expect that some combination of the coupling $C_1$ and parameter $\rho_h$ should correspond to the final equilibrated ratio (in some sense) of quark-gluon plasma and hadron gas in an effective way.

\vspace{.5cm}
\emph{Acknowledgement} - I would like to thank A. Davody and N. Abbassi for useful discussions.


\end{document}